\documentclass[%
reprint,
 amsmath,amssymb,
prl,
]{revtex4-2}

\usepackage{graphicx}
\usepackage{dcolumn}
\usepackage{bm}
\usepackage{xcolor}
\usepackage{epsfig}
\usepackage{bbold}
\usepackage{url}
\usepackage{subfig}
\usepackage{lineno}
\usepackage[colorlinks=true,citecolor={blue},urlcolor={blue}, linkcolor=blue]{hyperref}

\begin{document}

\title{
{Capturing many-body correlation effects with   quantum and classical computing }\\
}

\author{Karol Kowalski}
\email{karol.kowalski@pnnl.gov}
\affiliation{Physical Sciences Division, Pacific Northwest National Laboratory, Richland, Washington 99354, United States}

\author{Nicholas P. Bauman}
\affiliation{Physical Sciences Division, Pacific Northwest National Laboratory, Richland, Washington 99354, United States}

\author{Guang Hao Low}
\author{Martin Roetteler}
\affiliation{Azure Quantum, Microsoft, Redmond, Washington 98052, United States}

\author{John J. Rehr}
\author{Fernando D. Vila}
\email{fdv@uw.edu}
\affiliation{Department of Physics, University of Washington,
Seattle, Washington 98195, United States}

\date{\today}

\begin{abstract}
Theoretical descriptions of excited states of molecular systems in   high-energy regimes are crucial for supporting and driving many experimental efforts at light source facilities. However, capturing their complicated correlation effects requires formalisms that provide a hierarchical infrastructure of approximations. These approximations lead to an increased   overhead in classical computing methods, and therefore, decisions regarding the ranking of approximations and the quality of results must be made on purely numerical grounds. The emergence of quantum computing methods has the potential to change this situation. In this study, we demonstrate the efficiency of Quantum Phase Estimator (QPE) in identifying core-level states relevant to x-ray photoelectron spectroscopy. We compare and validate the QPE predictions with  exact diagonalization and real-time equation-of-motion coupled cluster formulations, which are some of the most accurate methods for states dominated by collective correlation effects.

\end{abstract}

\maketitle


\paragraph{Introduction.---}
Studies of the excited states of quantum systems corresponding to their complex excitation manifolds are crucial to advancing various scientific domains such as chemistry, physics, materials science, and biology. Advanced theoretical modeling tools can facilitate the understanding of various processes, including energy transfer through photochemical processes \cite{mcconnell2010energy,mirkovic2017light}, photocatalytic hydrogen production \cite{kudo2009heterogeneous,teets2011photocatalytic}, and carrier dynamics in nanoparticles and materials \cite{kongkanand2008quantum,hartland2011optical}. These tools are also needed to realize proton-coupled transfer in redox reactions and enable water oxidation \cite{yamaguchi2014regulating}, photoactivation processes in proteins
\cite{harper2003structural}, bioluminescence of living organisms \cite{tsien1998green}, and ultrafast protective mechanisms in DNA \cite{pecourt2001dna,sobolewski2004ab}. Predictive modeling tools also play a crucial role in supporting advanced light sources that  contribute significantly to the advancement of X-ray spectroscopies, including X-ray absorption, X-ray emission, resonant inelastic X-ray scattering, X-ray magnetic circular dichroism, and X-ray photoelectron, which have greatly improved our understanding of the structure and properties of matter
\cite{rehr2000theoretical,stohr2013nexafs,greczynski2020x}.


In this Letter, we investigate the practicality of  algorithms that leverage quantum or classical computational resources to describe high-energy excited states of ionized molecules in the context of X-ray photoelectron spectra (XPS) experiments. Specifically, we examine the Quantum Phase Estimation (QPE) algorithm for quantum computing.  To assess its accuracy, we compare with classical computing results obtained through exact diagonalization, or equivalently, full configuration interaction (FCI) methods and with systematic approximations based on recently developed the real-time equation-of-motion coupled cluster (RT-EOM-CC) method.


\paragraph{Quantum Phase Estimation.---} 
The Quantum phase estimation algorithm
\cite{nielsen2002quantum,kitaev1995quantum,Kitaev_1997,Abrams1999QPE,childs2010relationship,reiher2017elucidating}
allows one to estimate the eigenvalue $\lambda$ corresponding to an eigenvector $|\psi_\lambda\rangle$ of a general many-body Hamiltonian operator $H$,
i.e., $H |\psi_\lambda\rangle = \lambda |\psi_\lambda\rangle$.
From the QPE algorithm, the distribution of energies for the ground and excited states  is determined by the Hamiltonian $H$ and a trial many-body wavefunction $|\phi\rangle$ represented as a combinations of Slater determinants, wherein the probability of obtaining an energy estimate for a particular state is proportional to the amount of overlap of the trial wave with that corresponding eigenstate. %
Through repeated simulations, one accumulates samples from this distribution of eigenstate energies.
The error in each QPE energy estimate is inversely proportional to the number of applications of the time evolution operator $U=e^{-iHt}$, specified either through the number of ancillary qubits used in QPE or through the targeted bits-of-precision in the robust phase estimation variant that uses only one ancillary qubit.
%
In contrast to the variational quantum eigensolver (VQE) \cite{peruzzo2014variational,mcclean2016theory,romero2018strategies,Kandala2017,kandala2019,izmaylov2019unitary,lang2020unitary,grimsley2019adaptive,grimsley2019trotterized,mcardle2020quantum,Love2021,tilly2022variational}, which can only be used for energy estimates of a single targeted state (subject to the convergence of the iterative procedures), 
the QPE method can identify energies of states that have non-zero overlap with the trial wave function.
Furthermore, 
the design of VQE simulations requires \textit{a priori} knowledge of many-body effects needed to describe state of interest. 
Thus, the utilization of QPE techniques presents a unique prospect to identify complex states that cannot be readily obtained through traditional classical computing and   approximate methods. 

Several  quantum algorithms have been developed for the evaluation of Green's functions, which can be used in the calculation of ionization potential energies or as a solver for different embedding approaches
\cite{bauer2016hybrid,yoshimura2016measuring,endo2020calculation,kosugi2020construction,bassman2021simulating,baker2021lanczos,daniel2021sparse,sakurai2022hybrid,libbi2022effective,huggins2022nearly,keen2022hybrid,cao2023ab,gomes2023computing,dhawan2023quantum}.
In this Letter, we discuss a direct quantum computing approach to
evaluate the spectral function 
\begin{equation}
A(\omega)= -\frac{1}{\pi}\sum_p \mathrm{Im}(G^{\rm IP}_{pp}(\omega)), 
\label{eq0}
\end{equation}
where    $G_{pp}^{\rm IP}(\omega)$ corresponds to the diagonal elements of the ionization-potential part of the one-body Green's function (GF).
The $G_{pp}^{\rm IP}(\omega)$ can be obtained as a by-product of statistically averaged QPE simulations, i.e.,
\begin{eqnarray}
G_{pp}^{\rm IP}(\omega)&=&
\sum_i \frac{\langle\Psi_0^{(N)}|a_p^{\dagger}|\Psi_i^{(N-1)}\rangle\langle \Psi_i^{(N-1)}|a_p|\Psi_0^{(N)}\rangle}
{(\omega+(E_{i, FCI}^{(N-1)}-E_{0,FCI}^{(N)}) - i \theta)}  \label{eq1} \\
&\simeq& \sum_i \frac{P^{\rm QPE}(E_{i},|\theta_p\rangle)}{(\omega+(E_{i, QPE}^{(N-1)}-E_{0,QPE}^{(N)}) - i \theta)},
\label{eq2}
\end{eqnarray}
where $\theta$ is a broadening parameter, and $E_{0,QPE}^{(N)}$, $E_{i, QPE}^{(N-1)}$ are the ground state energy of $N$ electron system and the energies of $N-1$ electron systems, respectively. These are obtained   using trial states
\begin{equation}
|\theta_p\rangle = a_p |\Psi_0^{(N)}\rangle.
\label{eq3}
\end{equation}
Here $a_p^{\dagger}$/$a_p$ are creation/annihilation operators for an electron occupying the $p$-th spin-orbital.
If (1) QPE simulations are performed for all spin-orbitals of the $N-1$ electron system using $|\theta_p\rangle$ trial states, and (2) QPE is used to evaluate $E_{0,QPE}^{(N)}$, then one can reproduce, in an approximate way, the diagonal elements of the Green's function and corresponding spectral function.
For simplicity, we will also assume that $|\Psi_0^{(N)}\rangle$ can be approximated by the ground-state Hartree-Fock (HF) Slater determinant $|\Phi_0^{(N)}\rangle$, that is $|\theta_p\rangle \simeq a_p |\Phi_0^{(N)}\rangle$.
Within this assumption, we may always increase accuracy without changing the asymptotic cost by projecting $|\Phi_0^{(N)}\rangle$ or any other suitable reference onto $|\Psi_0^{(N)}\rangle$ by QPE. Subsequently, the quantum circuit that projects onto $a_p|\Psi_0^{(N)}\rangle$ always succeeds with probability $\frac{1}{2}$. \color{black} Therefore 
the Lehmann amplitudes in the numerator of
 Eq.\ (\ref{eq1}) can be approximated as 
\begin{equation}
    |\langle\Psi_i^{(N-1)}|a_p|\Psi_0^{(N)}\rangle|^2 \simeq 
    |\langle\Psi_i^{(N-1)}|\theta_p\rangle|^2 = P^{\rm QPE}(E_{i},|\theta_p\rangle),
    \label{eq4}
\end{equation}
where $P^{\rm QPE}(E_{i},|\theta_p\rangle)$ is the QPE probability of obtaining the state corresponding to the $E_i^{(N-1)}$ FCI energy (see Eq.\ (\ref{eq2})). We  illustrate the numerical efficiency of this approach 
in evaluations of  binding energies of inner electrons. For this energy regime we  limit the summation in Eq. (\ref{eq0}) to a single term
defined by the spin-orbital $c$, corresponding to the one-electron state for the core, i.e., 
\begin{equation}
A_c(\omega)\simeq -\frac{1}{\pi} \mathrm{Im}(G^{\rm IP}_{cc},(\omega)) 
\label{eq5}
\end{equation}
where
\begin{equation}
G_{cc}^{\rm IP}(\omega) \simeq
\sum_i \frac{P^{\rm QPE}(E_{i},|\theta_c\rangle)}{(\omega+(E_{i, QPE}^{(N-1)}-E_{0,QPE}^{(N)}) - i \theta)}.
\label{eq6}
\end{equation}
In our quantum computing simulations we employ the QPE implementation in the  Quantum Development Kit (QDK).\cite{low2019q,qdk}

\paragraph{Full configuration interactions.---} 
For comparison, our FCI  simulations were performed using the stringMB code, an occupation number representation-based emulator of quantum computing. In this code the action of the creation/annihilation operators for the electron in the $p$-th spin-orbital ($a_p/a_p^{\dagger}$) on the Slater determinants can be conveniently described using the occupation number representation, where each Slater determinant is represented as a vector
\begin{equation}
|\mathbf{n}\rangle = |n_M \; n_{M-1}\;  \ldots\;  n_{i+1}\; n_i \;n_{i-1} \;\ldots \;n_1 \rangle.
\label{onr}
\end{equation} 
and 
\begin{eqnarray}
a_i^{\dagger} |\mathbf{n}\rangle   &=& (-1)^{\sum_{k=1}^{i-1} n_k} 
|\mathbf{n}^{(+i)}\rangle \;\; ({\rm for}\; n_i=0)
\label{onr1} \;\;\;\;\;\; \\
a_i |\mathbf{n}\rangle  &=& (-1)^{\sum_{k=1}^{i-1} n_k} 
|\mathbf{n}^{(-i)}\rangle \;\; ({\rm for}\; n_i=1),
\label{onr2}
\end{eqnarray} 
where 
\begin{eqnarray}
|\mathbf{n}^{(+i)}\rangle &=& |n_M \; n_{M-1}\;  \ldots\;  n_{i+1}\; 1 \;n_{i-1} \;\ldots \;n_1 \rangle, \label{strp} \\
|\mathbf{n}^{(-i)}\rangle &=& |n_M \; n_{M-1}\;  \ldots\;  n_{i+1}\; 0 \;n_{i-1} \;\ldots \;n_1 \rangle. \label{strm} 
\end{eqnarray}
In the above equation, the occupation numbers $n_i$ are  either 1 (electron occupies $i$-th spin orbital) or 0 (the $i$-th spin orbital is empty). In Eq.\ (\ref{onr}), $M$ stands for the total number of spin-orbitals used to describe a quantum system, and $M=2n$, where $n$ is the number of orbitals.  
The stringMB code  allows us to construct a matrix representation (${\bf A}$) 
of general second-quantized operators $A$, where $A$ can be identified with the electronic Hamiltonian ($H$) or any function $f(H)$ of it.

\paragraph{Real-time Equation-of-Motion Coupled Cluster.---}
As an alternative based on classical computational methods, we have recently developed a real-time equation-of-motion coupled cluster (RT-EOM-CC) approach\cite{doi:10.1063/5.0004865,vila2021equation,vila2020real,vila2022real} to compute the core one-electron Green's function\cite{PhysRevB.90.085112,PhysRevB.91.121112,PhysRevB.94.035156,doi:10.1116/6.0001173} based on a CC form of the cumulant Green's function approximation. We found that the cumulant approximation produces accurate spectral functions for extended systems.\cite{PhysRevB.90.085112,PhysRevB.91.121112,PhysRevB.94.035156,doi:10.1116/6.0001173} 
Briefly, in the RT-EOM-CC method, the retarded GF is expressed as
\begin{equation}
\label{eq:cum_gf}
G_{c}^{R}(t) = -i \Theta(t) e^{-i (\epsilon_c + E_N^{corr}) t} e^{C_c^{R}(t)}.
\end{equation}
where $E_N^{corr}$ is the correlation energy of the $N$-electron ground state, as above $c$ corresponds to the spin-orbital associated with the excited hole, $\epsilon_c$ is the bare single-particle energy of this orbital, and $C_c^{R}(t)$ is its associated retarded cumulant.
The two main approximations in our implementation of RT-EOM-CC are the separable approximation $\left|0\right> \simeq a_c^\dagger \left|N-1\right>$, where $\left| N-1 \right>$ is the fully correlated $N-1$ electron
component of the exact $N$-electron ground state wave function $\left| 0 \right>$, and the use of a time-dependent (TD) CC ansatz for $\left| N-1 \right>$, i.e., $\left| N-1, t \right> = \tilde N(t) e^{T(t)} \left| \phi \right>$. Here $\tilde N(t)$ is a normalization factor and $T(t)$ is a TD CC operator. 
The $T(t)$ operator produces excted configurations in the $N-1$ electron space when acting on the reference determinant 
$\left| \phi \right>$ with a hole in level $c$, i.e., $\left| \phi \right> = a_c \left| \Phi_0^{(N)} \right>$. Here, as above, $\left| \Phi_0^{(N)} \right>$ is the Hartree--Fock Slater determinant of the $N$ electron ground state.
The cumulant in Eq. (\ref{eq:cum_gf}) is defined through its time derivative as a function of the time-dependent CC amplitudes:
\begin{equation}
\label{eq:dcdt}
\begin{split}
-i\frac{d C_c^R(t)}{dt} =& \sum_{ia} f_{ia} t_i^a(t)
 + \frac{1}{2} \sum_{ijab} v_{ij}^{ab} t_j^b(t) t_i^a(t)\\
 +& \frac{1}{4} \sum_{ijab} v_{ij}^{ab} t_{ij}^{ab}(t).
\end{split}
\end{equation}
Here, the $N-1$ electron Fock operator is defined as $f_{pq} = \epsilon_p \delta_{pq} - v_{pc}^{qc}$, $\epsilon_p$ is the energy of spin-orbital $p$, and we use antisymmetrized two-particle Coulomb integrals $v_{pq}^{rs} = \left< pq \left| \right| rs \right>$ over the generic spin-orbitals $p, q, r, s$.
The time-dependent amplitudes $t_{ij...}^{ab...}(t)$ in Eq. (\ref{eq:dcdt}) are determined by solving a set of coupled, first-order non-linear differential equations with initial conditions $t_{ij...}^{ab...}(0)=0$, which result in $C_c^R(0)=0$ for the cumulant in Eq. (\ref{eq:cum_gf}). These equations are analogous to those in static CCSD implementations.\cite{vila2022real}
In contrast to linearized self-energy-based formulations, Eq.\ (\ref{eq:dcdt}) shows that a CC ansatz results in a GF with a naturally explicit, non-perturbative exponential cumulant form.\cite{langreth70,SG1978,Hedin99review,sky}
We have previously demonstrated\cite{vila2022real} that the RT-EOM-CCSD method gives accurate coreand valence binding energies, with a mean absolute error (MAE) from experiment of $\sim$0.3 eV, and also provides a quantitative treatment of the many-body satellite region.

\paragraph{Computational Details.---} 
To assess the quality of the QPE and RT-EOM-CCSD predictions versus the FCI results for core-level spectral functions we compare results for a H$_2$O benchmark system described by the nine lowest (5 occupied and 4 virtual) restricted Hartree-Fock orbitals in the cc-pVDZ basis set.\cite{ccpvdz} Although the reduced size of the molecular basis precludes high-accuracy comparisons to the experimental XPS binding energies, we nevertheless compared the shifted RT-EOM-CCSD, QPE, and FCI results against the experimental results in order to assess their accuracy in the highly-correlated satellite region. The geometry of the water molecule corresponds to its 
equilibrium structure
with $R_{\rm OH}=0.9772\;$\AA $\;$ and $\angle{\rm HOH}=104.52^{\circ}$ \cite{bauman2020toward}.
The RT-EOM-CCSD simulation used Fock operator elements $f_{pq}$ and Coulomb $v_{pq}^{rs}$ integrals computed using the TCE\cite{TCE1,TCE2,TCE3,TCE4} implementation of RT-EOM-CCSD in NWChem.\cite{nwchem} The time integration of the equations-of-motion for the amplitudes used the 1st-order Adams-Moulton linear multi-step method described in Ref. \cite{doi:10.1021/acs.jctc.3c00045}, with a time step of 0.050 au (1.2 as) and a total simulation time of 900 au (22.5 fs). These parameters ensure that we achieve the resolution needed to compare to the FCI and QPE results.


\begin{figure}[t]
\includegraphics[clip,trim=0.5cm 0.0cm 0.5cm 0.0cm,width=0.50\textwidth]{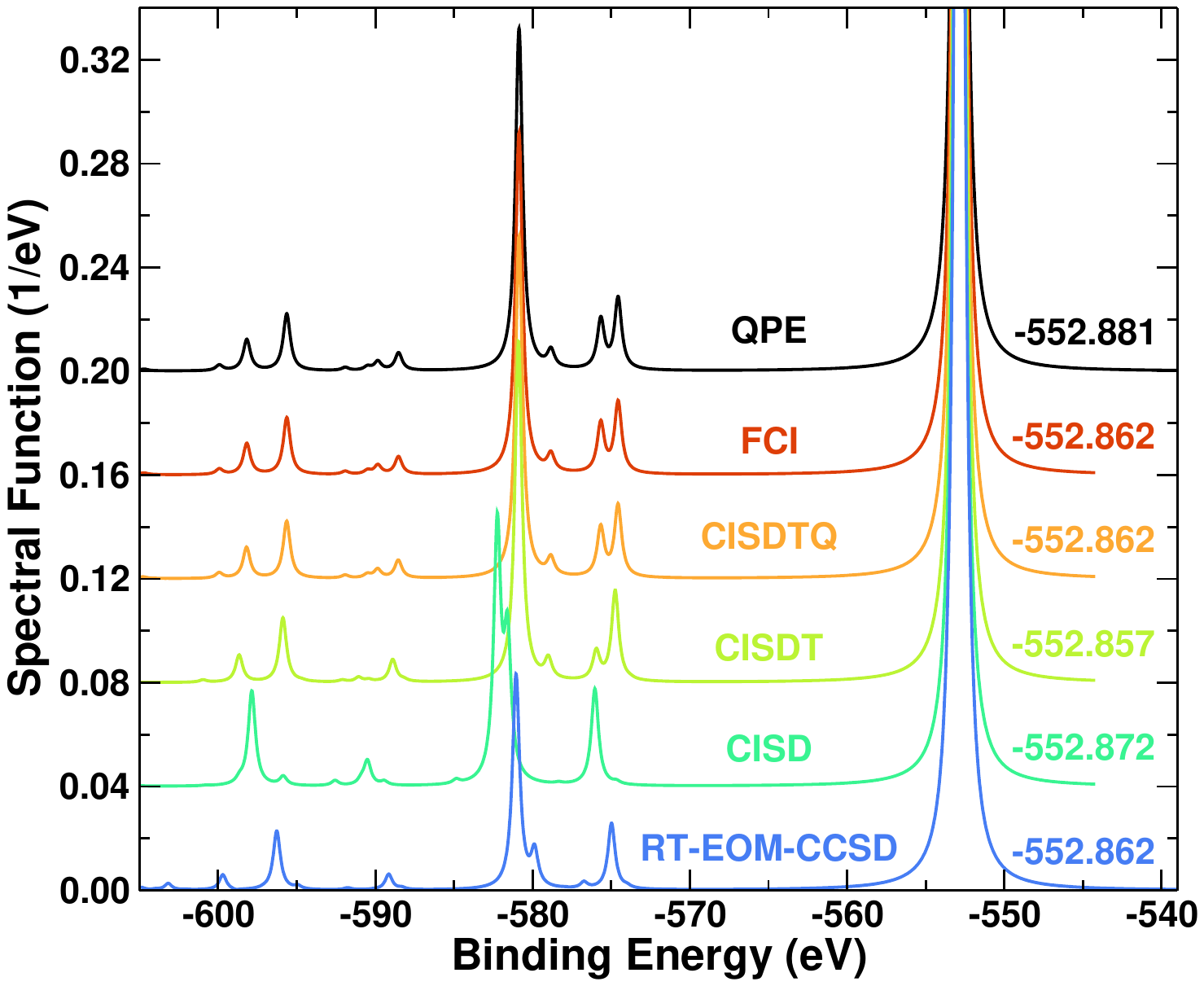}
\caption{\label{fig:SPFComp_Meth}
Comparison of the spectral functions of H$_2$O computed with the QPE, FCI, CISDTQ, CISDT, CISD, and RT-EOM-CCSD methods. The values next to the quasiparticle peak indicate its position for each method.
}
\end{figure}

\begin{figure}[t]
\includegraphics[clip,trim=0.5cm 1.0cm 0.5cm 2.8cm,width=0.50\textwidth]{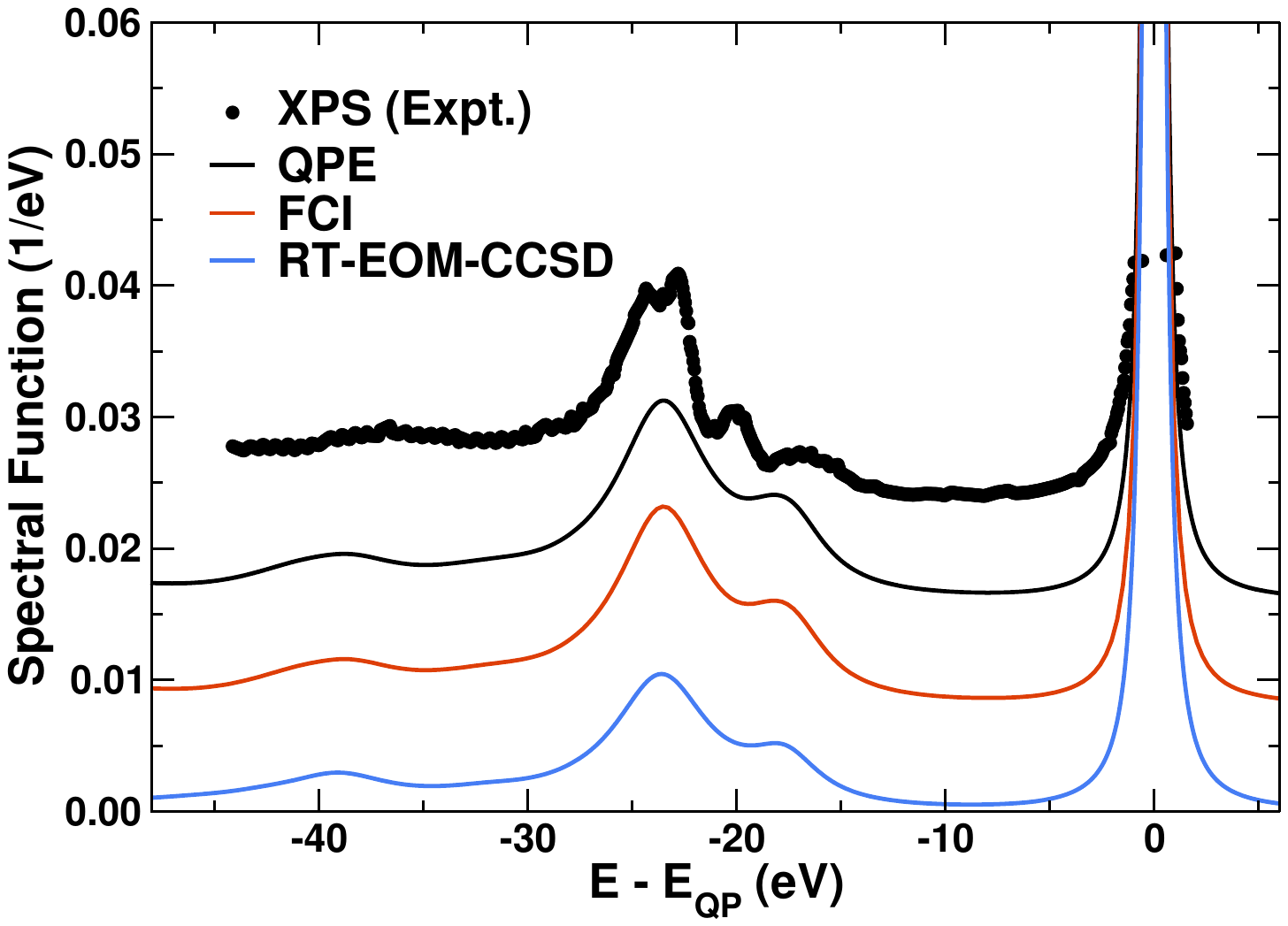}
\caption{\label{fig:SPFComp_Expt}
Comparison of the FCI, QPE and RT-EOM-CCSD broadened and scissors-shifted satellite regions of the spectral function of H$_2$O to the experimental XPS.\cite{SANKARI200651}
}
\end{figure}

Figure \ref{fig:SPFComp_Meth} shows a comparison of the 1a$_1^{-1}$ (O 1s) spectral functions of H$_2$O computed with the QPE, FCI and RT-EOM-CCSD approaches, as well as SD, SDT and SDTQ truncated CI results. 
The RT-EOM-CCSD results are in very good agreement with those from FCI and QPE. For instance, the quasiparticle peak is only 0.002 eV away from the exact FCI value. The agreement is also fairly good for the satellites, with the position of the main satellite at -580.86 eV being overestimated by only 0.2 eV. As shown in Table \ref{tab1},
the other satellites are also in overall good agreement. The only notable exception is the first satellite pair at -574.56 and -575.65 eV in the FCI, which appear as a single peak at -574.98 eV in the RT-EOM-CCSD.
Despite these minor differences, Fig. \ref{fig:SPFComp_Expt} shows that when the spectral functions are broadened and a scissors correction of 4.3 eV is applied to compare with experiment, their satellite weight distributions are nearly identical. Moreover, even with the small basis sets used in these calculations, the general theoretical satellite weight distribution is in reasonable agreement with the experiment.

In Tables \ref{tab1} and \ref{tab2}  QPE phases are collected corresponding to simulations employing two trial states $|\Phi_T(1)\rangle=|\Phi_1\rangle$ and $|\Phi_T(2)\rangle=|\Phi_{14}^6\rangle$ associated with the process of removal of electron from $1s$ orbital (orbital 1) ($|\Phi_T(1)\rangle$) followed by the valence type excitation of electron from the 4-th to the 6-th orbital ($|\Phi_T(2)\rangle$). The trial states
$|\Phi_T(1)\rangle$ and $|\Phi_T(2)\rangle$ correspond to singlet $A_1$ spatial symmetry. 
The QPE energies are evaluated through repeated simulations to accumulates samples from the distribution of eigenstate energies.
We used the QPE protocol as described in Ref.\ \cite{bauman2020toward}. For $|\Phi_T(1)\rangle$ and  $|\Phi_T(2)\rangle$ 
we collected $\simeq$ 500 and $\simeq$ 660 samples, respectively. Tables \ref{tab1} and \ref{tab2} show the averaged QPE energies. 
Using $|\Phi_T(1)\rangle$ as a trial state allows us to accurately identify the binding energy of the quasiparticle peak (FCI value of -552.86 eV). The probability of success (0.84) is also in very good agreement with its exact FCI counterpart ($|\langle\Psi^{\rm FCI}_i|\Phi_T(1)\rangle|^2$) of 0.82. 
This indicates that using QPE to generate the exact $|\theta_p\rangle$ would have succeeded with high probability, in which case $P^{\rm QPE}=P^{\rm FCI}$.\color{black}
The RT-EOM-CCSD results also show very good agreement with the FCI beyond the quasiparticle, with a maximum error of 0.6 eV and an average deviation of just 0.3 eV.
The high accuracy of the QPE predictions is also illustrated by the values of binding energies corresponding to other states with non-zero overlap with $|\Phi_T(1)\rangle$. 
We note that the QPE protocol may always produce  predictions within an arbitrarily smaller targeted error $\epsilon$ of FCI with only an extra multiplicative $\sim1/\epsilon$ cost.\color{black}  The use of the $|\Phi_T(2)\rangle$ trial state enables us to explore other classes of core-level binding effects: e.g., corresponding to more complex states (compared to the $|\Phi_T(1)\rangle$ Slater determinant), dominated by double excitations with respect to the $|\Phi_0^{(N)}\rangle$ $N$-electron state. The QPE simulations with 
$|\Phi_T(1)\rangle$ and $|\Phi_T(2)\rangle$ can also identify the same binding energy (FCI energy of -574.56 eV in Table \ref{tab2}) with the probability of success in the QPE simulations being proportional to the overlap between trial and exact wave functions.
Figures \ref{fig:SPFComp_Meth} and \ref{fig:SPFComp_Expt} show excellent agreement of the QPE spectral function with the FCI one, and qualitative agreement with the experimental data.

\renewcommand{\tabcolsep}{0.2cm}
\begin{table}
    \centering
       \caption{Comparison of the highest probability $P^{\rm QPE}$ QPE averaged binding energies (in eV) with the RT-EOM-CCSD (labelled CCSD) and FCI results for the 9-orbital cc-pVDZ model of the H$_2$O system. The QPE results were obtained using the $|\Phi_T(1)\rangle$ trial Slater determinant corresponding to the ${}^2A_1$ symmetry. The exact ($P^{\rm FCI}$) and RT-EOM-CCSD ($P^{\rm CCSD}$) probabilities were obtained with the FCI and RT-EOM-CCSD codes. For example, in the FCI case, the probability is equal to the square of the coefficient corresponding to the  $|\Phi_T(1)\rangle$  determinant.}
    \begin{tabular}{c c c c c c}
    \hline \hline \\[-0.2cm]
  QPE & $P^{\rm QPE}$ &CCSD & $P^{\rm CCSD}$ &FCI & $P^{\rm FCI}$ \\
  \hline \\[-0.2cm]
   -595.77   &  0.02 &  -596.27 & 0.02 &-595.63 & 0.01\\[0.1cm]
   -580.87   &  0.07 &  -581.05 & 0.07 &-580.86 & 0.08\\[0.1cm]
   -574.72   &  0.01 &  -574.98 & 0.02 &-574.56 & 0.01\\[0.1cm]
   -552.81   &  0.84 &  -552.86 & 0.82 &-552.86 & 0.82\\[0.1cm]
         \hline \hline
    \end{tabular}
    \label{tab1}
\end{table}

\renewcommand{\tabcolsep}{0.2cm}
\begin{table}
    \centering
       \caption{Comparison of the highest probability $P^{\rm QPE}$ QPE averaged binding energies (in eV) with the FCI results for the 9-orbital cc-pVDZ model of the H$_2$O system. The QPE results were obtained using the $|\Phi_T(2)\rangle$ trial Slater determinant corresponding to the ${}^2A_1$ symmetry.
 The exact probabilities $P^{\rm FCI}$ where obtained with the FCI code and are equal to the square of the coefficient corresponding to the $|\Phi_T(2)\rangle$ determinant in the FCI wave function expansion.}
    \begin{tabular}{c c c c}
    \hline \hline \\[-0.2cm]
   QPE & $P^{\rm QPE}$ &  FCI & $P^{\rm FCI}$ \\
\hline \\[-0.1cm]
   -588.43 & 0.08  & -588.52 & 0.06 \\[0.1cm]
   -587.70 & 0.02  & -587.96 &  0.04 \\[0.1cm]
   -574.51 & 0.44  & -574.56 & 0.42\\[0.1cm]
   -573.58 & 0.22  & -573.65 & 0.22 \\[0.1cm]
         \hline \hline
    \end{tabular}
    \label{tab2}
\end{table}

Based on the analysis of the RT-EOM-CCSD results, it can be inferred that this time-dependent variant of CCSD can accurately describe the energies of multiple ionized states. 
This property can be attributed to the unique nature of real-time CC approximations compared to their stationary counterparts. The Linked Cluster Theorem (LCT)
\cite{goldstone1957derivation,brandow1967linked}
demonstrates that the low-order quadruple excitation in the configuration interaction expansion for the ground state can be approximated by low-order doubly excited cluster amplitude products.
However, the situation is different in the time domain, especially in terms of the phases of time-dependent CC amplitudes. For instance, using the simplest  partitioning of the Hamiltonian where the unperturbed part is defined by the diagonal part of the Fock operator (with $\epsilon_p$ being the orbital energies), the products of the 0-th order time-dependent singly excited amplitudes
$
t_i^a(\tau)^{(0)} = t_{i}^{a}(0)
e^{i(\epsilon_a-\epsilon_i)\tau}
$
can replicate the phases of zeroth-order approximations to doubly ($t_{ij}^{ab}(\tau)^{(0)}$), triply ($t_{ijk}^{abc}(\tau)^{(0)}$), etc., excited amplitudes:
$
t_{ij}^{ab}(\tau)^{(0)} =
t_{ij}^{ab}(0)e^{i(\epsilon_a+\epsilon_b-\epsilon_i-\epsilon_j)\tau} $
and
$
t_{ijk}^{abc}(\tau)^{(0)} = 
t_{ijk}^{abc}(0)e^{i(\epsilon_a+\epsilon_b+\epsilon_c-\epsilon_i-\epsilon_j-\epsilon_k)\tau}$.
This suggests that in the real-time case, various rank cluster amplitudes are correlated in a different way compared to the stationary situation. In the above expressions, indices $i,j,k,\ldots$ ($a,b,c,\ldots$) designate occupied (unoccupied) spin-orbitals in $|\Phi_0^{(N)}\rangle$.
The effectiveness of the RT-EOM-CC formalism in representing multiple electronic states using a single CC Ansatz may be attributed mainly to the phase additivity. While stationary non-linear CC equations are characterized by multiple solutions \cite{kowalski1998towards}, their accuracy beyond the ground-state solution, which is consistent with the LCT, is often less pronounced compared to the time-dependent case. This observation is demonstrated in Figure \ref{fig:SPFComp_Meth}, where the RT-EOM-CCSD spectral function is compared with CISD, CISDT, CISDTQ, and FCI counterparts revealing feature characteristic for high-order CI approximations (CISDT and CISDTQ). 

\paragraph{Summary.---} 
In this Letter, we investigated the efficacy of QPE and RT-EOM-CCSD formulations for the evaluation of spectral functions of ionized states in high-energy regimes. To assess their effectiveness, we compared these formulations to exact FCI results. Our findings demonstrate that the approximate QPE-derived spectral function can reproduce all features of the exact spectral function in the analyzed XPS binding energies energy window, including both main and satellite peaks. The information required to construct the Lehmann  representation of the spectral function in quantum simulations is a by-product of statistically averaged QPE simulations. Similarly RT-EOM-CCSD simulations using a  generalization of the static CCSD Ansatz gave   spectral functions that faithfully reproduce the features of the QPE and FCI results. This behavior can be attributed to the additive separations of the phases of cluster amplitudes in the lowest order of perturbation theory, as discussed earlier. Overall, our results show that both QPE and RT-EOM-CCSD and formulations can accurately evaluate the exact spectral functions of ionized states in high-energy regimes.  These findings open new pathways for treatments of many-body effects in complex systems that classical computing algorthms cannot handle, and  have potential applications in various fields, including materials science, chemistry, and physics.

This material is based upon work supported by Quantum Science Center (QSC), a National Quantum Information Science Research Center of the U.S. Department of Energy (under FWP  76213).
This work was also supported by the Computational Chemical Sciences Program of the U.S. Department of Energy, Office of Science, BES, Chemical Sciences, Geosciences and Biosciences Division in the Center for Scalable
and Predictive methods for Excitations and Correlated phenomena (SPEC) at Pacific Northwest National Laboratory under FWP 70942.
With computational support from NERSC, a DOE Office of Science User Facility, under contract no. DE-AC02-05CH11231.

\bibliography{QPE_FCI_EOM_Main}

\end{document}